# GPLAN: Computer-Generated Dimensioned Floorplans for given Adjacencies

Krishnendra Shekhawat, Nitant Upasani, Sumit Bisht, Rahil Jain

Department of Mathematics, BITS Pilani, Pilani Campus, India-333031

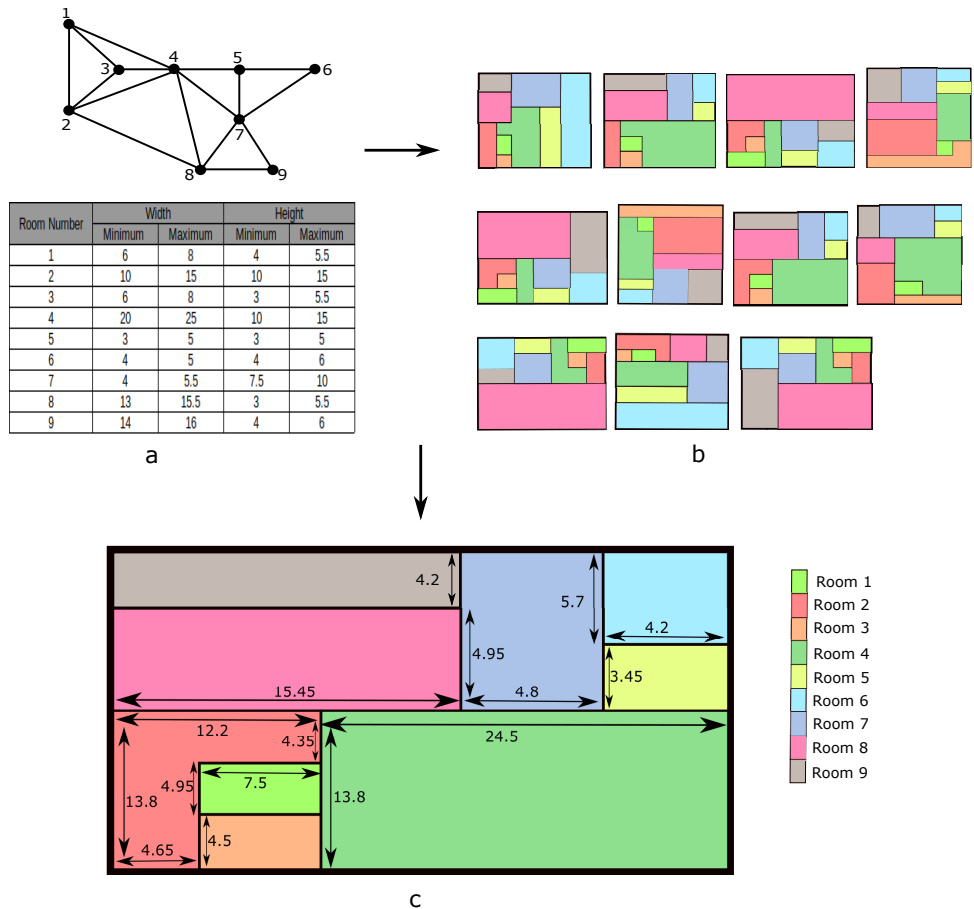

**Fig. 1.** The working of GPLAN: Illustrating the input format of GPLAN and output produced by it a) Input adjacency graph and dimensional constraints b) Multiple dimensionless floorplans satisfying all the adjacency requirements c) A dimensioned floorplan obtained from the dimensionless floorplans satisfying all given dimensional constraints

**Abstract.** In this paper, we present GPLAN, software aimed at constructing dimensioned floorplan layouts based on graph-theoretical and optimization techniques. GPLAN takes user requirements as input in the following two forms:

i. Adjacency graph: It allows user to draw an adjacency graph on a GUI (graphical user interface) corresponding to which GPLAN produces a set of dimensioned floorplans with a rectangular boundary, where each floorplan is topologically distinct from others.
ii. Dimensionless layout: Here, user can draw any layout with rectangular or non-rectangular boundary on a GUI and GPLAN transforms it into a dimensioned floorplan while preserving adjacencies, positions, shapes of the rooms.

The above approaches represent different ways of inserting adjacencies and GPLAN generate dimensioned floorplans corresponding to the given adjacencies. The larger aim is to provide alternative platforms to user for producing dimensioned floorplans for all given (architectural) constraints, which can be further refined by architects.

## 1 Introduction

From 1970s, a lot of research has been done in the domain of computer-aided architectural design, where the prime focus is to automatically generate floorplan layouts, so that these layouts are regarded as initial layouts by architects/designers and can be further modified and adjusted by them. After a lot of developments in this direction, there still lies a huge gap between the proposed research and its practical aspects. One of the major reason for this gap is the interdisciplinary nature of the floorplanning problem, i.e., the problem cannot be handled with a specific approach which may be based on architecture, mathematics, artificial intelligence, machine learning, computer science, etc. The larger aim of this work is to bridge this gap by considering each constraint one by one and by identifying and applying a specific technique based on the nature of a sub-problem. In this paper, we attempt to handle those sub-problems which are mathematical in nature.

### 1.1 Literature Review

The automated generation of architectural layouts using graph theory began with the generation of rectangular floorplans (RFP). The first attempt in this direction was made by Levin [1] in the early 1960s. Then in the coming years, many researchers proposed graph-theoretical approach for the enumeration or construction of rectangular layouts [2,3]. In 1975, the first computer algorithm was given by Sauda [4] for enumerating all topologically distinct RFPs having up to eight rooms. In the 1980s, comprehensive studies were presented for the existence and construction of a rectangular dual (dimensionless RFP) whose prime focus was VLSI design [5–8]. During this time, Roth et al. [9] and Rinsma [10,11] developed efficient graph algorithms for the construction of dimensionless and dimensioned RFPs with some restrictions on the input graph.

During the early 1990s, researchers realised that there are graphs for which RFPs do not exist [12,13]. In 1995, Giffin et al. [14] gave a linear time algorithm for constructing an orthogonal floorplan (OFP) for a given planar triangulated graph (PTG) with module area requirements. Using the concept of orderly spanning trees, in 2003, Liao et al. [15] gave a linear time algorithm for constructing a dimensionless OFP for any $n$-vertex PTG which require fewer module types, i.e., the algorithm uses only $I$-modules, $L$-modules and $T$-modules, but $Z$-modules could not be incorporated. In 2010, Marson et al. [16] restricted their work to sliceable floorplans and generated layouts having aspect ratios close to one, without considering the adjacency constraints. In 2011, Jokar and Sangchooli [17] used the concept of face area of a graph for the construction of dimensionless OFPs for given PTGs. In 2012, Eppstein et al. [18] gave a method for finding an area universal RFP[1] for the given adjacency requirements whenever such layout exists. They also gave the necessary and sufficient condition for an RFP to be area universal, i.e., an RFP is area universal if and only if it is one-sided. In 2013, Alam et al. [19] gave the construction of an area-universal OFP for a given PTG. In 2018, Wang et al. [20] presented the automated regeneration of well-known existing dimensionless RFPs while considering underlying adjacency graph of the existing floorplan. The proposed prototype is called GADG (graph approach to floor plan generation). In the same year, Shekhawat [21] enumerated all possible maximal RFPs without considering dimensions of the rooms. In 2020, Upasani et al. [22] developed a prototype for generating dimensioned RFPs for any drawn rectangular arrangement, satisfying adjacency, size and symmetric requirements. They do not take adjacency graph as an input. More recently, Wang and Zhang [23] extended GADG [20] for generating dimensioned OFPs corresponding to user-specified design requirements.

Shape grammar can be seen as a parallel and an efficient approach for automation where the idea is to generate designs through the execution of shape rules [24]. In 1995, Harada et al. [25] presented an interactive model for generating floorplans using shape grammars. In 2003, Wonka et al. [26] introduced split grammars for incorporating flexible design requirements to model buildings using a variety of styles. In 2005, Duarte [27] implemented a shape-grammar based technique to recreate Alvaro Siza's designs. Recently a lot of work has been done to generate building models and their 3D representations by extending the use of split grammars [28–30].

In the recent times, many new approaches have evolved for supervised automation in floorplan design. In 2010, Merell et al. [31] introduced a supervised learning algorithm (based on bayesian networks) for generating residential floor-plans. In this direction, in 2019, Wu et al. [32] also used a data-driven technique for constructing interior layouts with fixed outer boundaries. As an extension, Hu et al. [33] presented the generation of floorplans using a graph neural network (GNN) while considering room adjacencies in the form of layout graphs (layout graphs enable human users to provide sparse design constraints). This work is trained on the data-set provided by [32].

---

[1] an RFP is area universal if any assignment of areas to rectangles can be realized by a rectangular module.

As an alternative approach, in 2013, Rodrigues et al. [34] presented an evolutionary strategy for incorporating complex topological and geometric user requirements while successfully generating a feasible layout solution, but the implementation of the proposed work has not been discussed. In the same year, Bao et al. [35] constructed good floorplan layouts using simulated annealing, which are characterised by parameters like lighting, heating and circulations. In 2018, Wu et al. [36] presented a hierarchical framework and used mixed integer quadratic programming for the dimensioning of layouts with fixed exterior boundaries.

Considering the automation where user can insert his choices, in 2019, Nisztuk et al. [37] built a tool for automated floorplan generation, covering the majority of adjacency and size constraints, but it is limited to rectangular rooms only and generates empty spaces in the layouts. Recently, Shi et al. [38] used reinforcement learning based on a heuristic search technique called Monte Carlo Tree Search to generate a closest feasible dimensionless RFP corresponding to any adjacency graph inserted by the user.

Because of the stochastic nature of algorithms presented in the above papers, their time-complexity is very high and are thus not suitable for complex building design. Also, most of the above-discussed work is restricted to a single layout for the given constraints. Clearly, multiple layouts allow the possibility to building practitioners for analysing, comparing these designs and choosing the most appropriate one. In 2012, Regateiro et al. [39] produced multiple dimensioned floorplans without considering adjacency relation using block algebra, but are unable to generate all possible solutions. Also, the proposed work does not give details about its implementation. In 2018, Zawidzki [40] first generated a set of candidate layouts based on adjacency constraints using a depth-first backtracking search algorithm and then included other customised objectives based on user satisfaction to give an optimal architectural layout. However, it took eight hours to generate 30 candidate solutions only. Nisztuk et al. [37] also produced multiple solutions using a greedy approach and thus have very high computational time which clearly shows the efficiency of graph algorithms over greedy search techniques.

## 1.2 Gaps in the existing literature and proposed work

The gaps in the existing literature can be listed as follows:

i. Dimensionless floorplans (with rectangular boundary): Corresponding to given adjacencies, there exist linear time algorithms for constructing dimensionless rectangular or orthogonal floorplans [8, 13, 15, 38, 41], but we did not find a computer-based tool that can generate a floorplan for any given planar triangulated adjacency graph. In this work, we present a software GPLAN that provides a GUI to the user for drawing an adjacency graph, and then generates a floorplan layout while satisfying all the adjacency requirements. It first prefers to generate a rectangular layout if it exists; otherwise, an orthogonal layout is generated. It is also possible to check the existence of an RFP corresponding to given graph using GPLAN.

ii. Dimensioned floorplans: For a given adjacency graph, there are algorithms for generating dimensioned RFP [9, 11] and dimensioned OFP [23] but a generalized optimization technique for producing a feasible floorplan for any given dimensions of the rooms is not available. GPLAN provides a GUI where user can insert dimensions of all the rooms and using optimization techniques, it produces dimensioned floorplans satisfying given adjacencies as well as dimensions.

iii. Multiple floorplans: GPLAN is capable of generating all possible topologically distinct floorplans corresponding to given adjacency constraints, which is not common in the existing literature. In particular, the generation of multiple OFPs has not been done previously.

iv. Time complexity: A lot of existing work is capable of producing residential building layouts because they have a small number of rooms [31, 36, 37], but for the complex building structures with a large number of rooms, we need efficient algorithms. GPLAN generates a variety of layouts in a few seconds.

v. Irregular floorplans (IFP): Irregular floorplans are floorplans with non-rectangular boundary (see Figure 2c). In the recent times, some work has been done for building IFPs for given adjacencies [32], but it is limited to a small number of rooms, and proposed algorithms cannot be generalised to any adjacency graph. In particular, there exist efficient algorithms for constructing dimensionless RFPs and OFPs corresponding to given adjacency relations, but there does not exist such an algorithm for IFPs. Therefore, to produce dimensioned IFP, GPLAN generates a GUI where a user can draw any dimensionless IFP and can give dimensional constraints as input. It then produces a dimensioned IFP while preserving adjacency relations of underlying dimensionless IFP along with the positions and shapes of the rooms.

vi. Re-generation of floorplans: There exists a limited work for the re-generation of floorplans [20, 42]. A well-known architectural floorplan $F$ (RFP or OFP) can be re-generated using GPLAN by extracting the underlying adjacency graph $G$ of $F$ and then by generating a floorplan corresponding to $G$. In this case, GPLAN also produces floorplans which are topologically distinct to $F$. A well-known IFP can also be reproduced by drawing it on a GUI, but GPLAN does not generate topologically distinct IFPs.

## 2 Preliminaries

In this section, we present a few important terminologies which are used frequently in literature and also throughout this paper.

**Definition 1.** *Floorplans.* A floorplan is a partition of a finite-sized polygon $P$ into a finite set of dimensioned polygons $\{P_1, P_2 \ldots P_n\}$ called rooms. An irregular floorplan (IFP) has non-overlapping rooms with no restrictions on outer boundary $P$. Orthogonal floorplans (OFPs) are a particular case of IFPs where $P$ is a rectangle. Moreover, when the contained polygons $P_1, P_2 \ldots P_n$ are all internally disjoint rectangles, and the envelope $P$ is convex, the floorplan is said to be a rectangular floorplan (RFP).

Two rooms in a floor plan are adjacent if they share a wall or a section of it, where a wall of a room refers to the edges forming its perimeter.

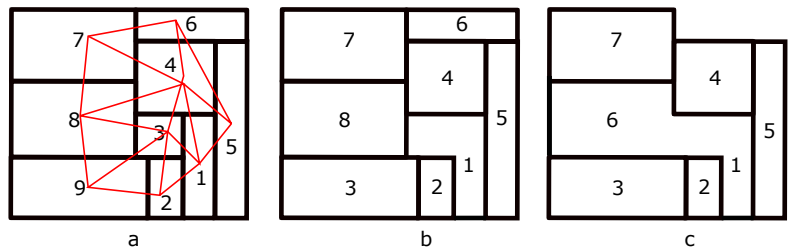

**Fig. 2.** Floorplans Typology a) Rectangular floorplan (RFP) b) Orthogonal floorplan (OFP) c) Irregular floorplan (IFP)

**Definition 2.** *Graphs.* A graph is a set of vertices and edges denoted by $\mathcal{G}(n, m)$, where $n$ and $m$ denote the number of vertices and edges respectively. A graph is said to be planar if it can be embedded in the plane without crossing of edges; otherwise, it is a non-planar graph (the graph in Figure 3 is planar while the graph in Figure 4a is non-planar). A plane graph is a planar graph with an embedding that divides the plane into connected components called faces/regions (the graph in Figure 3 has 20 internal faces and 1 exterior face).

In architectural terms, an adjacency graph is a graph that provides a specific neighbourhood between the given rooms. For each floorplan, there exists a graph known as the weak dual graph, which can be constructed by replacing each room with a vertex and adding an edge to the vertices that correspond to the adjacent rooms (see Figure 2a where red edges show the weak dual graph of the floorplan).

**Definition 3.** *Separating Triangle.* A separating triangle a-b-c is a cycle of length three in a graph $G$ such that $G$ - $\{a, b, c\}$ is disconnected. For example, in Figure 4c, the cycle 1-5-3 is a separating triangle because the graph becomes disconnected on the removal of vertex 6.

**Definition 4.** *Properly Triangulated Planar Graph (PTPG)* [5], [7] A connected planar graph is triangulated if all of its faces (except the exterior) are triangular; exterior face can or cannot be a triangle. This graph is called planar triangulated graph (PTG). The graph in Figure 4c is a PTG. A PTG with no separating triangle and with exterior face of the length at least 4 is called properly triangular planar graph (PTPG). For example, the adjacency graph shown in Figure 3 is a PTPG, whereas the graphs shown in Figure 4 are not PTPG.

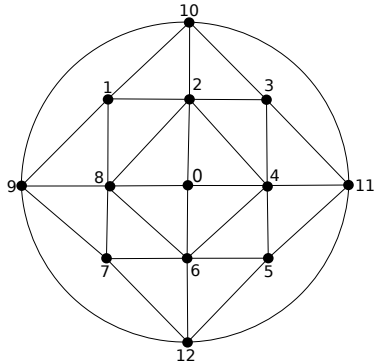

**Fig. 3.** A properly triangulated planar graph (PTPG)

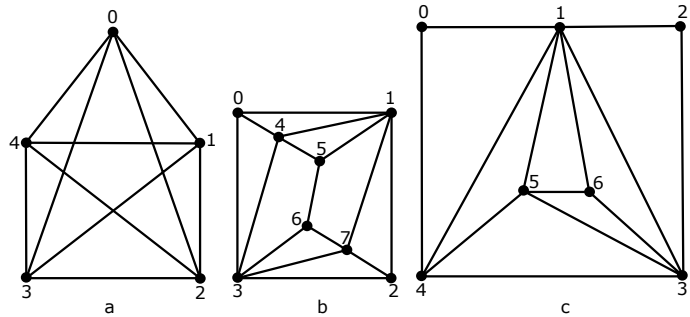

**Fig. 4.** Graphs that are not PTPGs a) A non-planar graph b) A non-triangulated graph c) A PTG with a separating triangle ($\triangle 153$)

**Definition 5.** *Regular Edge Labelling (REL) [41]* A regular edge labelling of a PTPG $G$ having the exterior face of length 4 is a partition of the interior edges of $G$ into two subsets $T_1$, $T_2$ of directed edges such that for each interior vertex $u$, the edges incident to $u$ appear in a counterclockwise order around $u$ as follows: a set of edges in $T_1$ leaving $u$, a set of edges in $T_2$ entering $u$, a set of edges in $T_1$ entering $u$ and a set of edges in $T_2$ leaving $u$.

Let $N, E, S, W$ be the four exterior vertices in a clockwise order. All interior edges incident to $N$ are in $T_1$ and entering $N$. All interior edges incident to $E$ are in $T_2$ and entering $E$. All interior edges incident to $S$ are in $T_1$ and leaving $S$. All interior edges incident to $W$ are in $T_2$ and leaving $W$. For example, a REL for the PTPG in Figure 3 is shown in Figure 5.

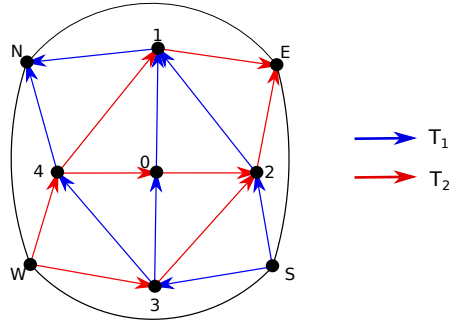

**Fig. 5.** Regular edge labelling

**Definition 6.** *Shortcut [7].* A graph $G$ is said to be bi-connected if after deleting any vertex of $G$, it remains connected, i.e., $G$ has no cut vertices. The graph in Figure 6a is 1-connected having vertices 3 and 4 as cut-vertices, while the graph in Figure 6b is bi-connected.

A shortcut in a planar bi-connected graph $G$ is an edge that is incident to two vertices on the outer boundary of $G$ but is not part of the outer boundary. For example, in Figure 6b, 0-1-2-3-4-5 forms the outer boundary of the graph and edges (1,5) and (2,4) are shortcuts.

**Definition 7.** *Corner implying Path (CIP) [7].* A corner implying path (CIP) in a planar bi-connected graph $G$ is a path $u_1, u_2, \ldots, u_n$ on the outer boundary of graph $G$ with the property that $(u_1, u_n)$ is a shortcut and $u_2, u_3, \ldots, u_{n-1}$ are not the endpoints of any shortcut. For example, in Figure 6b, 1-0-5 is a CIP because edge (1,5) is a shortcut and 0 is not an endpoint of any shortcut.

## 3 GPLAN for the construction of dimensioned floorplans corresponding to a given graph

This section talks about the working of GPLAN which has been developed in Python for constructing dimensioned floorplans for the given adjacency relations. In the coming subsections, we will talk about the existence and construction of floorplans (RFP and OFP) corresponding to a given adjacency graph. In Section 4, we discuss dimensioned irregular floorplans.

For GPLAN, the input for adjacencies can be given in the following two ways (refer to Figure 7a):

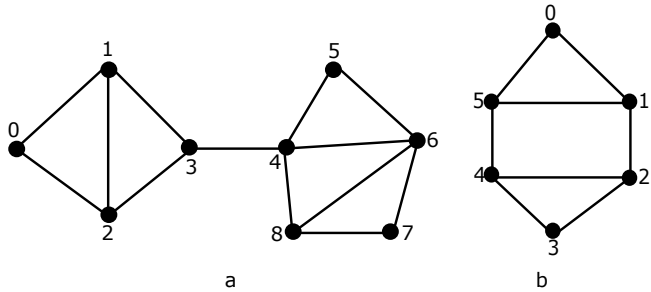

**Fig. 6.** Biconnectivity, Shortcut and Corner Implying Paths

1. In the form of an adjacency graph, as shown in Figure 7b (construction of floorplans (RFP and OFP) corresponding to given adjacencies is discussed in Section 3),
2. In the form of a dimensionless layout, as shown in Figure 7c (construction of dimensioned IFP corresponding to given adjacencies is discussed in Section 4).

Other than the adjacency constraints, GPLAN takes dimensional constraints as input, which is shown in Figure 7d.

### 3.1 Existence of a floorplan

A floorplan can be seen as a planar graph whose weak dual graph (see Definition 2) is always a PTG (see Definition 4), i.e., for a floorplan to exist corresponding to a given adjacency graph, it must be connected, planar and triangulated. In the case of a 4-joint in a floorplan, the weak dual graph has a cycle of length 4, but 4-joint is a limiting case of 3-joint as shown in Figure 8. Hence we have restricted to floorplans having only 3-joints. Since, a weak dual graph is always a PTG, the input graph for GPLAN must be a PTG. In GPLAN, if user inserts a non-planar or a non-triangular graph as an input, then it generates an error as shown in Figure 9.

A PTG can be 1-connected or bi-connected. It can be seen in the literature that the construction of floorplans exists for bi-connected PTGs only ( [8, 9, 15, 17, 20]) because they may be easy to handle and floorplans corresponding to bi-connected PTGs are comparatively architecturally significant. For a comparison, refer to Figure 10.

Hence, we are considering bi-connected PTG as an input for GPLAN, i.e., it generates an error if the given graph is not bi-connected PTG as shown in Figure 11.

*Remark 1.* If a graph is non-planar or non-triangulated, using the existing algorithms, it can be made planar [43] (by deleting a few edges) and triangulation [44] (by adding a few edges). A 1-connected PTG can be made bi-connected using the biconnectivity algorithm given in [45]. The following algorithm will be incorporated into GPLAN in the near future.

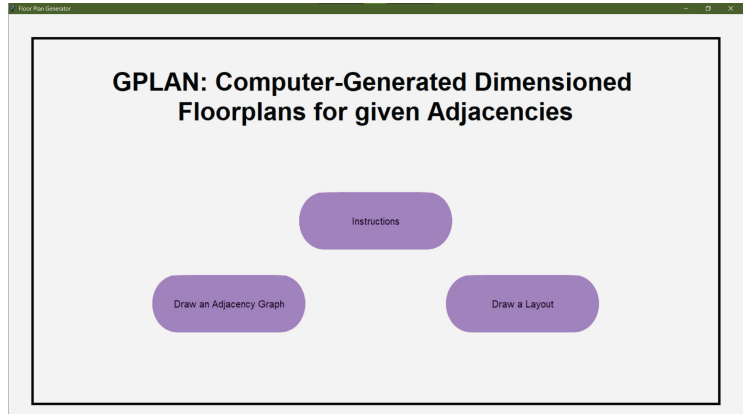
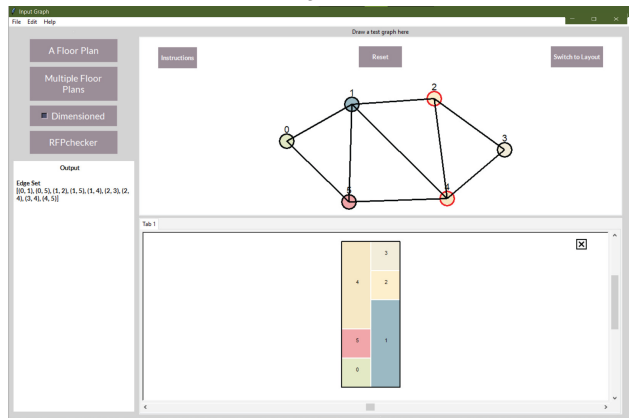
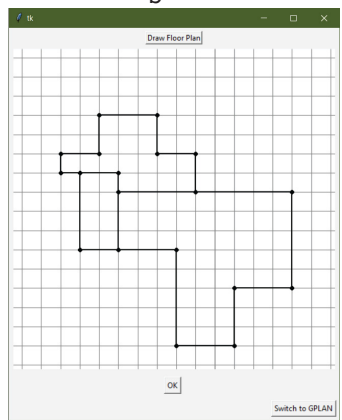
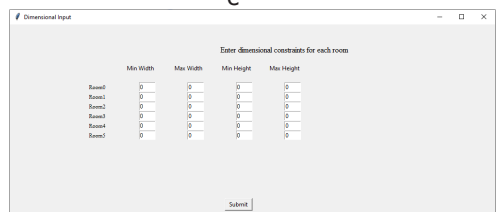

**Fig. 7.** GPLAN interface and input constraints

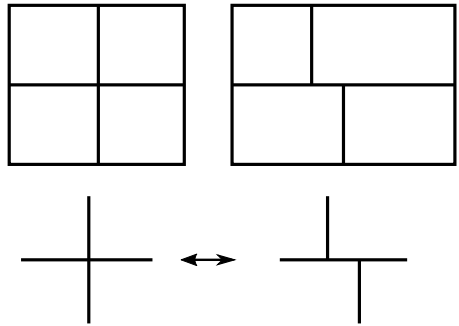

**Fig. 8.** 4-joint as a limiting case of 3-joint

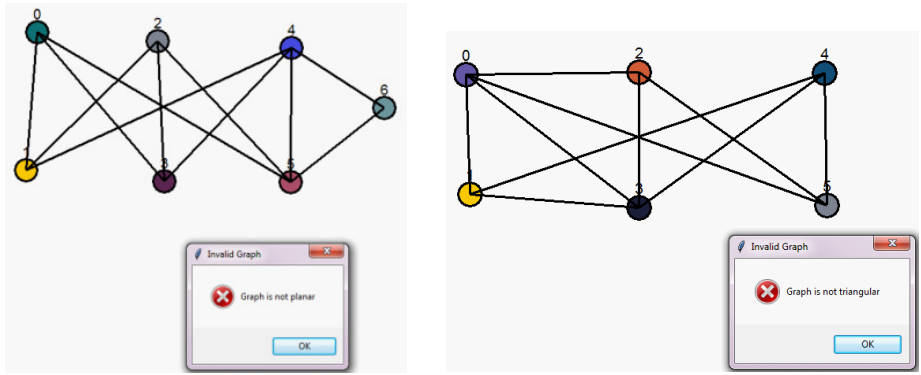

**Fig. 9.** GPLAN gave an error if the input graph is non-planar or not triangulated

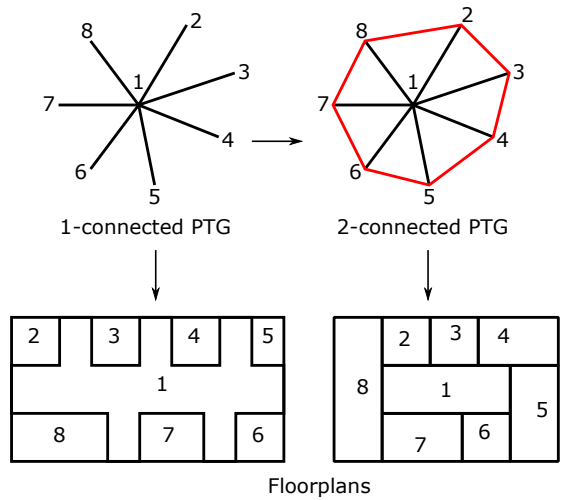

**Fig. 10.** 1-connected and bi-connected PTGs and corresponding floorplans

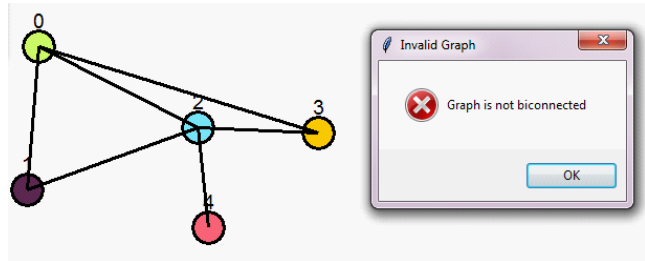

**Fig. 11.** 1-connected PTG

### 3.2 Existence of a rectangular floorplan (RFP)

Since most of the buildings are rectangular [46], we first prefer to construct a RFP for the given adjacency graph. If RFP does not exist, an OFP is constructed.

A dimensionless RFP is known as a rectangular dual, which only exist for adjacency graphs that are PTPGs (see Definition 4). In 1985, the following theorem [6] was proposed.

**Theorem 1.** A bi-connected PTPG $G$ has a rectangular dual if and only if it has no more than four corner implying paths (CIPs).

It is clear from Theorem 1 that checking the number of CIPs for a graph having a large number of vertices is not an easy calculation by hand. Hence, we incorporated **RFPchecker** to GPLAN so that user can draw any bi-connected PTG on the GUI and can check if there exists an RFP for the required graph. It also specifies the reason for the non-existence of an RFP. For an illustration, refer to Figure 12.

### 3.3 Construction of rectangular floorplans

From Figure 2a, we can see that the construction of a weak dual graph from its floorplan is very easy, but the converse is not true, i.e., for a given PTG, constructing its corresponding floorplan automatically requires an efficient algorithm and its implementation. We first discuss the construction of an RFP and then move to the construction of an OFP.

To generate RFPs for bi-connected PTPGs, we extend the rectangular dual finding algorithm proposed by [41] by defining the process of 4-completion using corner implying paths (CIPs).

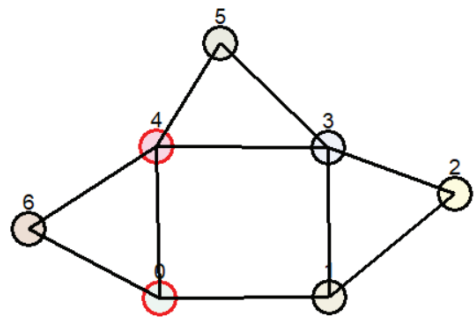
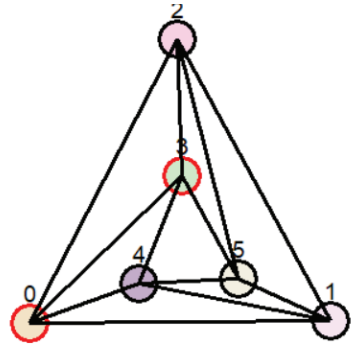
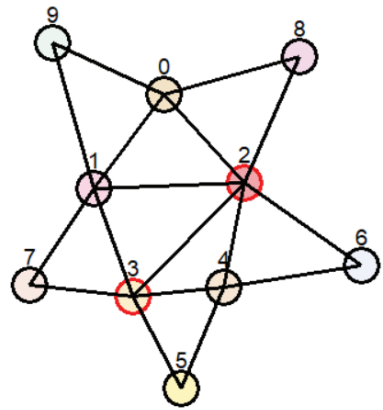

**Fig. 12.** Using RFPchecker, illustrating the cases for which RFPs do not exist

Kant and He [41] propose transforming a bi-connected PTG into a PTPG (satisfying Theorem 1) by adding four new vertices $N, E, S, W$ and connecting them to the exterior boundary of the input graph. To transform a bi-connected PTG to a PTPG, we select four vertices $u_0, u_1, u_2, u_3$ on the exterior face in a clockwise order. Let $P_i (i = 0, 1, 2, 3)$ be the path on the exterior face between $u_i$ and $u_{i+1}$ (if $i > 3$, then reduce $i$ by 4, i.e., $u_4$ is same as $u_0$). We connect $N, E, S$ and $W$ to every vertex in $P_0, P_1, P_2$ and $P_3$ respectively and add four new edges $(N, W), (N, E), (S, E)$ and $(S, W)$ to have the required PTPG (see Figure 13b).

The vertices $u_i (i = 0, 1, 2, 3)$ are selected in such a way that the addition of new edges doesn't lead to the formation of separating triangles. This can be done by choosing $u_i (i = 0, 1, 2, 3)$ as corner vertices which are obtained using the definition of CIP, and using the method proposed in [7]. Let the number of CIPs be $k(k \leq 4)$. We can pick any vertex from the interior of each of these $k$ paths and pick additional $4 - k$ vertices from the outer boundary to obtain the four corner vertices. Figure 13a shows an input graph $G$ and its CIPs. We choose vertex $1, 3, 5, 7$ from each CIP as corner vertices and then add extra vertices $N, E, W$ and $S$ to form a PTPG as shown in Figure 13.

Kant and He [41] do not consider the special case when the input graph is a triangle. For this case, we can choose 3 outer edges as $P_1, P_2, P_3$ and $P_4$ consisting of only one vertex which can be chosen arbitrarily. Figure 14 shows the 4-completion for a triangle with vertex 1 chosen as a path $P_4$.

Once we have a bi-connected PTPG, using the algorithms given in [41], we can construct the corresponding RFP. The major steps involved in the construction of a RFP are illustrated in Figure 15. GPLAN automatically generates a RFP corresponding to any bi-connected PTPG as illustrated in Figure 16.

*Remark 2.* At this stage, we are not considering the functionality of given spaces, but the user has a choice to insert the room names based on their functions as illustrated in Figure 16.

### 3.4 Construction of orthogonal floorplans

It is clear from Theorem 1 that there does not exist a RFP corresponding to a bi-connected planar triangulated graph $G$ if any of the following holds:

i. $G$ has more than four corner implying paths (see Figure 17a),
ii. $G$ has separating triangles (see Figure 17b),
iii. The exterior face of $G$ is triangular (see Figure 17c).

In all these cases, we need to add extra vertices to $G$ and then triangulate $G$ so that the modified $G$ has at most four corner implying paths, has no separating triangles and has exterior face non-triangular. In this case, we obtain an RFP for modified $G$ and then merge the rooms corresponding to the newly added vertices to build an OFP for $G$. For a better understanding refer to Figure 18 where the graph in Figure 18a has a separating triangle $\triangle 124$. In Figure 18b, an extra vertex 6 has been introduced for the removal of the separating triangle, and a new edge has been added to triangulate the modified graph. 18c represents an RFP corresponding to the graph in 18b. Now, in 18c, room 6 is the extra room which needs to be merged with either room 1 or room 2 to have an OFP as shown in Figure 18d.

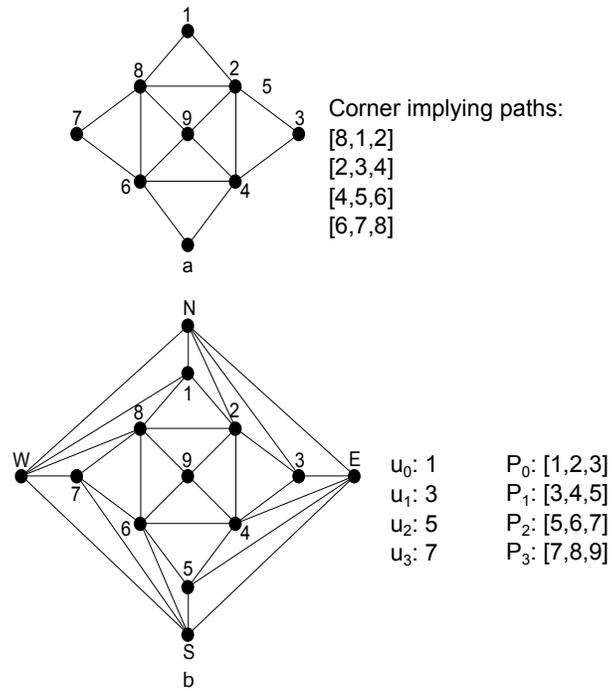

**Fig. 13.** a) A PTG with corner implying paths b) A PTPG derived from given PTG by using 4-completion considering $u_i (i = 0, 1, 2, 3)$ as corner vertices

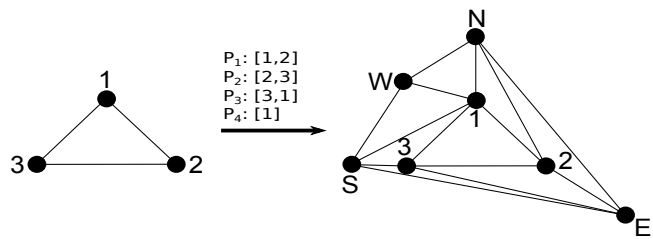

**Fig. 14.** 4-completion of a triangle

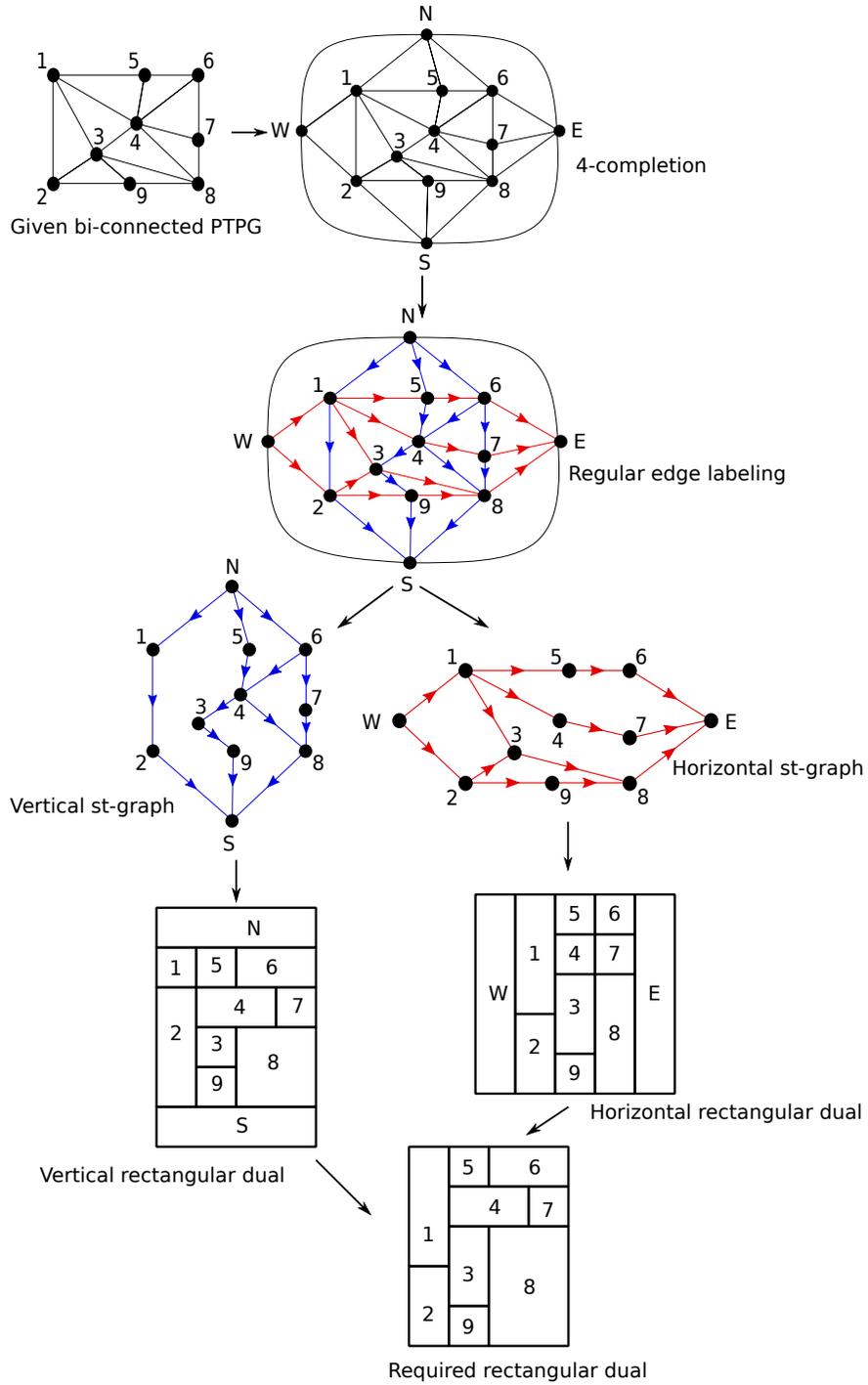

**Fig. 15.** Construction of a RFP corresponding to a given bi-connected PTPG

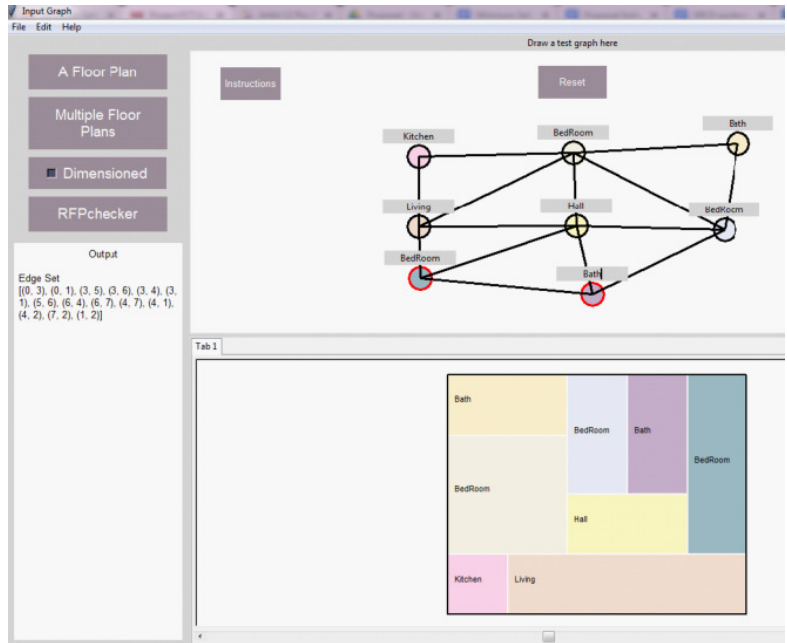

**Fig. 16.** An RFP generated using GPLAN

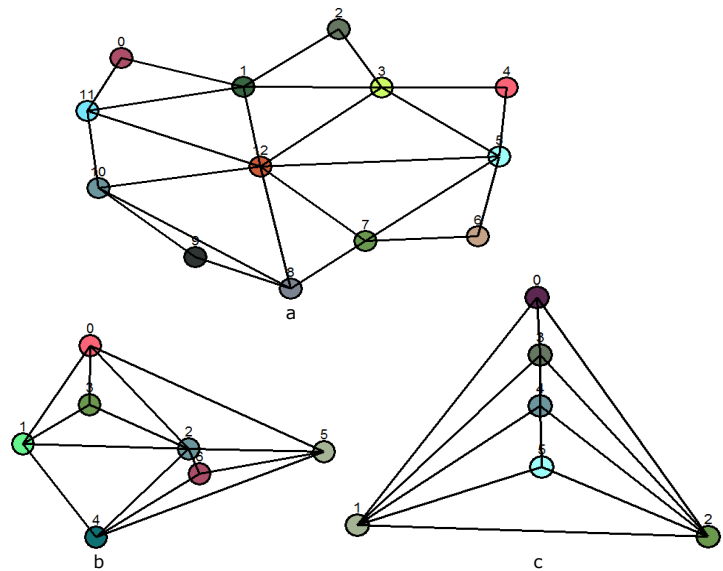

**Fig. 17.** Input graphs for which RFPs do not exist

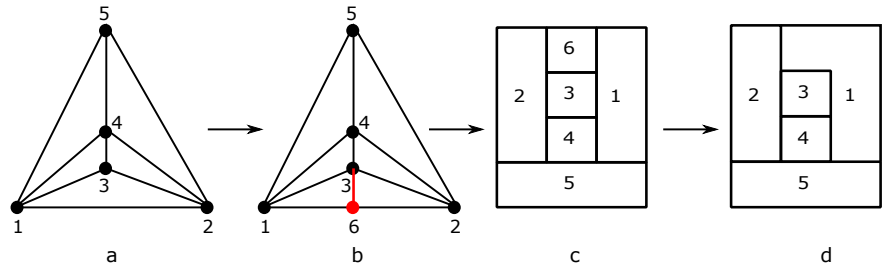

**Fig. 18.** Construction of an OFP corresponding to a given PTG

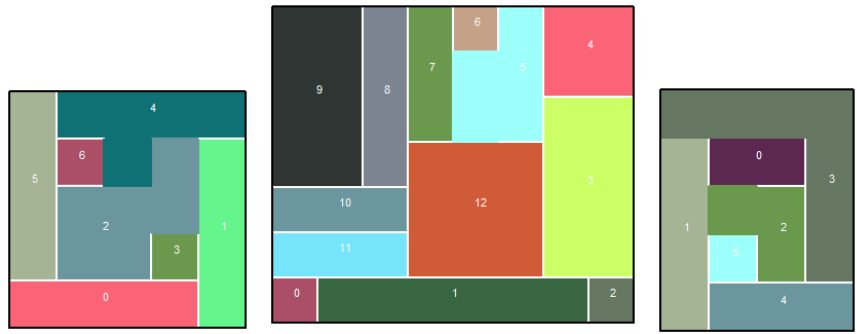

**Fig. 19.** OFPs for the graphs in Figure 17

The steps for the construction of an OFP for a bi-connected PTG have been incorporated into GPLAN. For an illustration, OFPs corresponding to each graph in Figure 17 are shown in Figure 19.

### 3.5 Multiple floorplans

Here, the idea is to construct all topologically distinct floorplans corresponding to a given PTG. Two floorplans are topologically distinct if they have the same underlying weak dual graph, but their horizontal and vertical adjacencies are different. For example, refer to Figure 20 where two topologically distinct floorplans are illustrated.

For a given graph, GPLAN iterates over different possible boundary paths and finds different RELs possible for that boundary path using the concept of flippable item [18] and hence generates all possible RFPs for the obtained PTPG. For the input graph in Figure 15, GPLAN iterates over all possible 154 boundaries and generates 1300 topologically distinct RFPs in 282.45 seconds. A few of topologically distinct RFPs generated using GPLAN are illustrated in Figure 21.

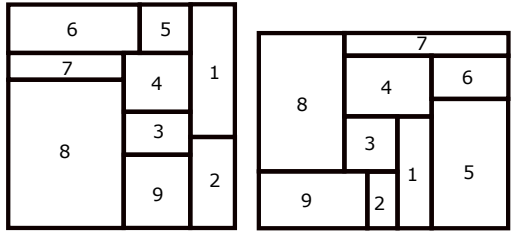

**Fig. 20.** Topologically distinct RFPs

For a bi-connected PTG $G$, GPLAN finds all possible ways to add extra vertices to $G$ so that $G$ has at most four corner implying paths (CIPs) and has no separating triangles (STs). For a graph with $k > 4$ CIPs, there can be $^kC_4$ ways to add extra vertices so that the graph has at most 4 CIPs. Similarly, there can be three possible ways for the removal of a ST from the graph. In this way, GPLAN finds all possible ways to convert a bi-connected PTG to a bi-connected PTPG for which an RFP exists. For each PTPG thus obtained, it generates all possible RFPs using the method described in Section 3.3 and for each RFP, it then merges the extra room to obtain all possible OFPs. For the input graph in Figure 17b, GPLAN generates 256 topologically distinct OFPs in 26.56 seconds, a few of which are illustrated in Figure 22.

### 3.6 Dimensioned floorplans

For incorporating the dimensional constraints into each topological solution, we implement an algorithm which is based on the network flow model proposed by Upasani et. al [22]. This algorithm is based on an iterative linear optimisation framework employed on the horizontal and vertical st-graphs (obtained during the construction of an RFP, Figure 15). However, in this paper, we propose an improved objective function which further minimises the total area of the RFP in comparison to [22]. Against the total width/height optimised in [22], we use the difference between width/height and their maximum bounds as the objective function, which has given more optimised results in terms of area of the floorplan. The optimisation problem for both the st-graphs is stated as follows:

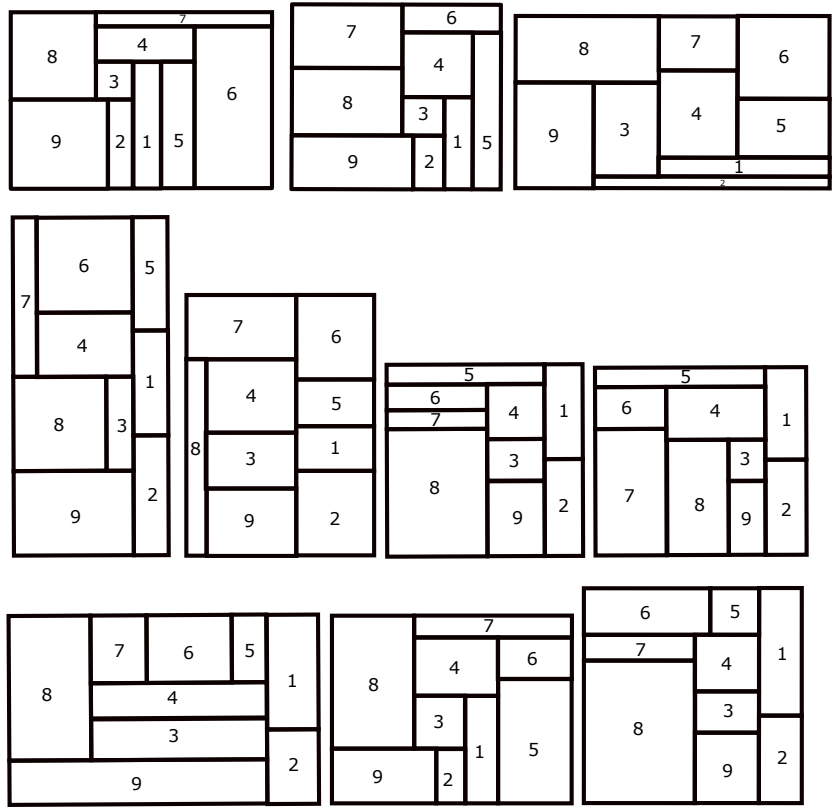

**Fig. 21.** Out of 1300 solutions, which can be obtained using GPLAN, a few RFPs corresponding to the input graph in Figure 15

$$Minimize: \quad \sum w(e_{j,i}) - d_i^{max}$$

$$such\ that: \quad \sum w(e_{ji}) = \sum w(e_{ik}) \quad \forall i \in V(G) \quad (1)$$

$$\min(d_i) \leq \sum w(e_{ji}) \leq \max(d_i) \quad \forall i \in V(G)$$

where $d_i^{max}$ is the maximum dimension (width/height) of room $i$, $\sum w(e_{ji})$ denotes the total inflow and $\sum w(e_{ik})$ denotes the total outflow from vertex $i$.

Figure 23 enlists all equality and inequality constraints, along with the objective functions associated with these st-graphs. Conforming to these constraints, GPLAN optimises width and height separately using the dual-simplex method to generate a feasible dimensioned floorplan. Higher efficiency of the simplex method, when compared to any other stochastic optimisation algorithms, helps to incorporate dimensions in all topological solutions within a reasonable time.

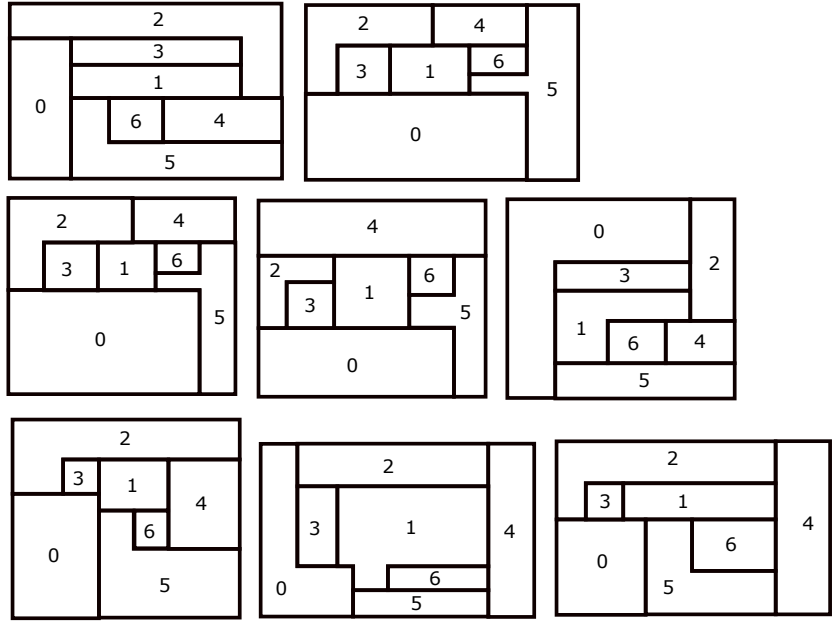

**Fig. 22.** Out of 256 solutions, which can be obtained using GPLAN, a few OFPs corresponding to the input graph in Figure 17b

Taking dimensional requirements as input is challenging for orthogonal rooms as the constructs of width and height are difficult to define in this case (see Fig. 24b). Thus, for dimensioning, we consider RFP before the merging of extra rooms (as show in Figure 24c). Customizing dimensions for rectangular partitions of an orthogonal room is convenient, as all such parts are simply integrated after network flow optimisation to yield a dimensioned OFP. A dimensioned OFP generated using GPLAN is shown in Figure 24d.

## 4 Dimensioned irregular floorplans

An irregular floorplan (IFP) has been shown in Figure 25a. Construction of an IFP for an adjacency graph is more challenging than the construction of an RFP or an OFP because, in both RFP and OFP, the boundary of the floorplan is fixed, i.e., rectangular, but in case of an IFP, the boundary of the layout is variable. In the literature, there does not exist any algorithm that talks about the existence and construction of an IFP corresponding to a given PTG. Hence, in this work, we take a dimensionless IFP as an input, which can be drawn on a GUI, and then generate a dimensioned IFP while satisfying the dimensions given by the user and preserving the adjacencies and topology of the rooms (which are also given by the user in terms of a dimensionless IFP).

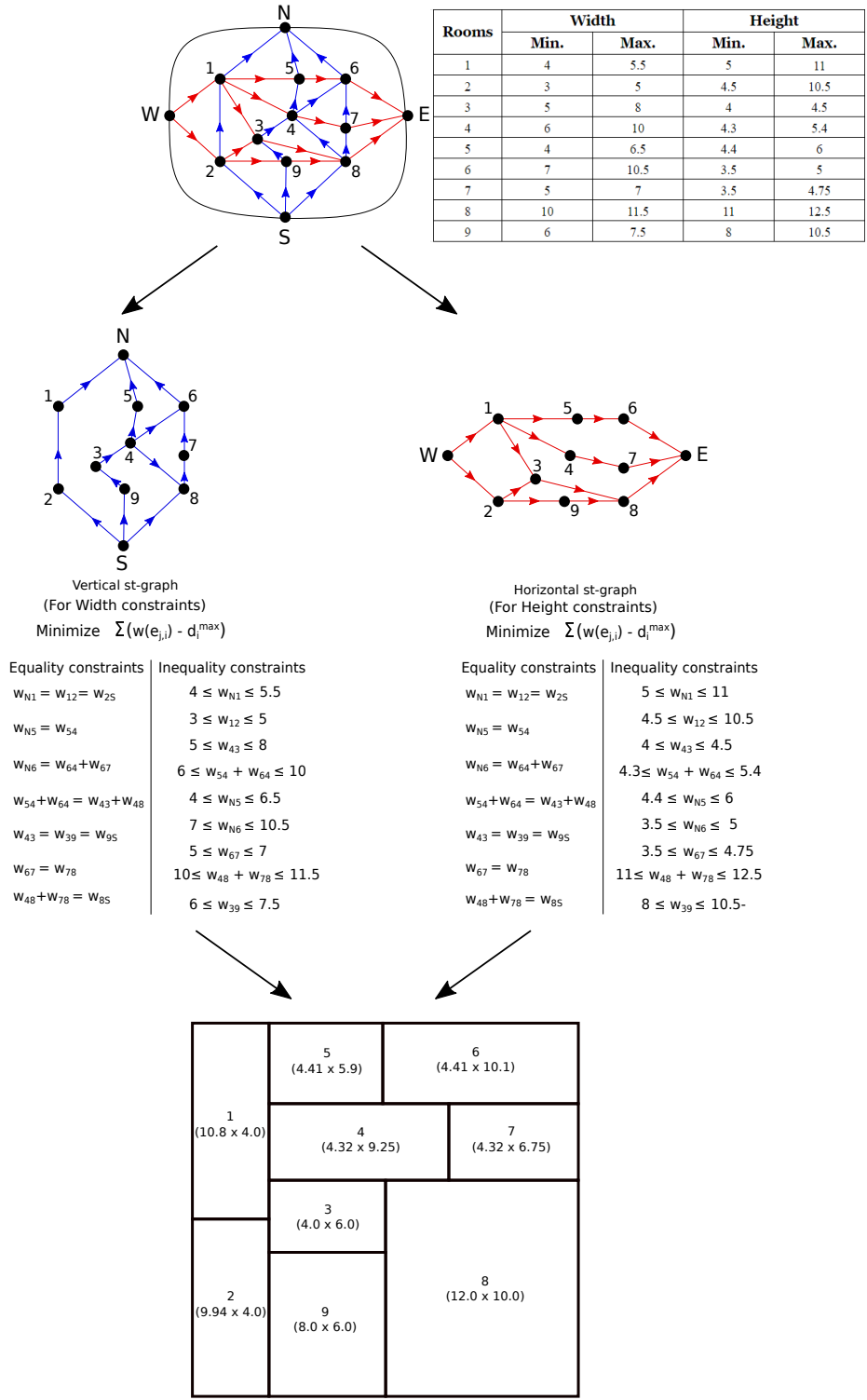

**Fig. 23.** Dimensioning model based on network flow and linear optimization

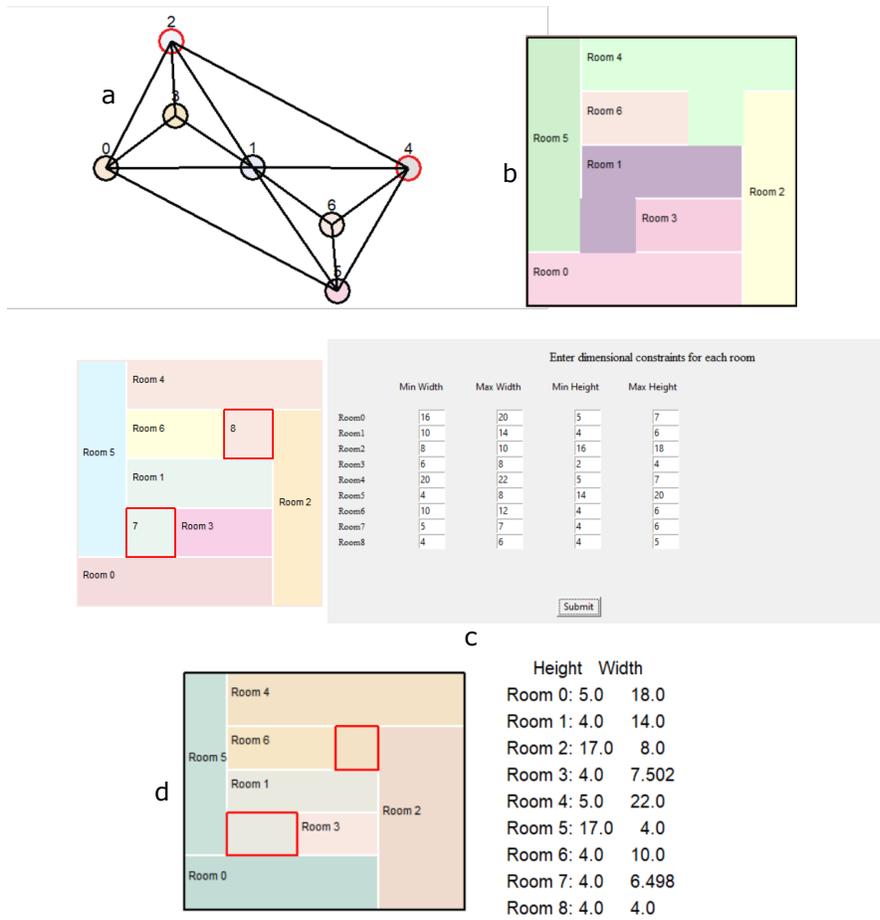

**Fig. 24.** a) A PTG which is not a PTPG b) Dimensionless OFP generated by GPLAN corresponding to given PTG c) Partitioned OFP and dimensional constraints d) Dimensioned OFP generated by GPLAN

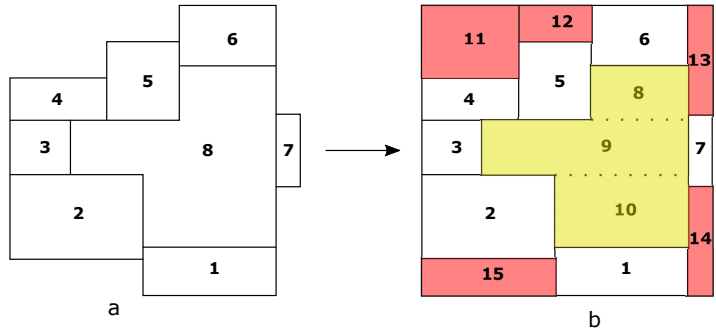

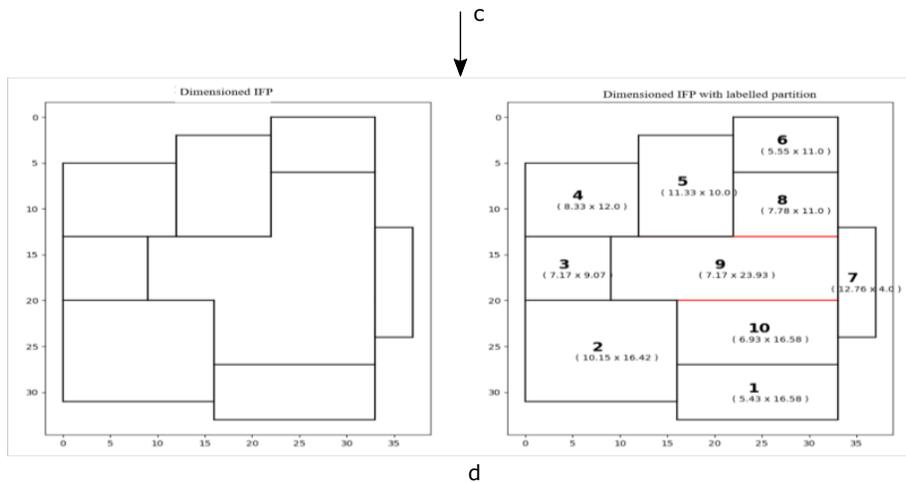

**Fig. 25.** Construction of a dimensioned irregular floorplan a) Dimensionless IFP as input b) IFP transformed into a RFP by adding extra rooms (red) and partitioning orthogonal rooms into rectangles (yellow) c) Dimensional requirements d) Resulting dimensioned IFP obtained using GPLAN

The input IFP may not always contain rectangular rooms; hence we partition the orthogonal rooms into a minimum number of rectangles, using [47]. However, to employ the network flow formulation for dimensioning, the exterior boundary of floorplan should also be rectangular. Therefore, extra rooms are added to dimensionless IFP, as shown in Figure 25b, such that the outer boundary assumes the shape of a rectangle. The dimensions of IFP are obtained similar to an RFP, i.e., by using the network flow optimisation [22]. However, no dimensional constraints are imposed on these additional rooms, as they are merely intermediaries, and will be removed from the final dimensioned floorplan, as illustrated in Figure 25d.

Since RFPs and OFPs are special cases of IFPs, the dimensioned RFPs and OFPs can also be generated by drawing corresponding dimensionless RFPs and OFPs on the GUI generated by GPLAN.

## 5  A Case Study

In this paper, we presented the automated generation of dimensioned floorplans with rectangular and non-rectangular boundaries, where rectangular boundary floorplans are generated corresponding to a given adjacency graph and floorplans having variable boundaries are produced based on the initial layout drawn by the user. It is exciting to see that both the approaches can be used for re-generating well-known existing architectural floorplans.

In the first case, we need to extract the underlying graph of the existing floorplan $F$ and then using GPLAN, we can re-generate $F$ while introducing the dimensions given by the user. For an illustration, refer to Villa Trissino floorplan in Figure 26a, designed by Scamozzi [48] in 1778. Its underlying graph is shown in Figure 26b and the re-generated Villa Trissino floorplan by GPLAN is demonstrated in Figure 26c. GPLAN efficiently produces floorplans that are topologically distinct to Villa Trissino floorplan, some of which are shown in Figure 26d. Clearly, GPLAN is capable of re-generating any floorplan with rectangular boundary and it also generates topologically distinct floorplans, which provides a set of alternatives to the existing floorplans.

In the second case, user can draw the existing floorplan on a GUI and GPLAN generates a dimensioned floorplan corresponding to the dimensional constraints. In this case, the user has more flexibility in choosing the floorplan he wants to re-generate, but alternative floorplans are not possible to generate. Here, the idea is to re-generate complex building structures with new dimensions because with time, the dimensional requirements change a lot. The steps of re-generation of Banstead Home School Plan (shown in Figure 27a) are shown in Figure 27, where Figure 27b is drawn by the user and is transformed into Figure 27c using GPLAN, after which dimensional constraints are provided by the user. Figure 27d presents the dimensioned Banstead Home School Plan re-generated using GPLAN.

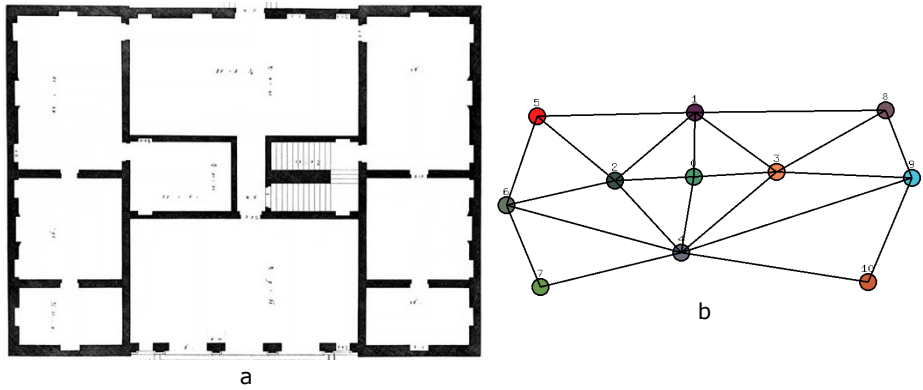
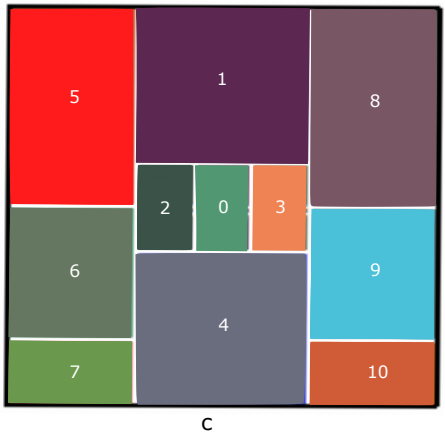
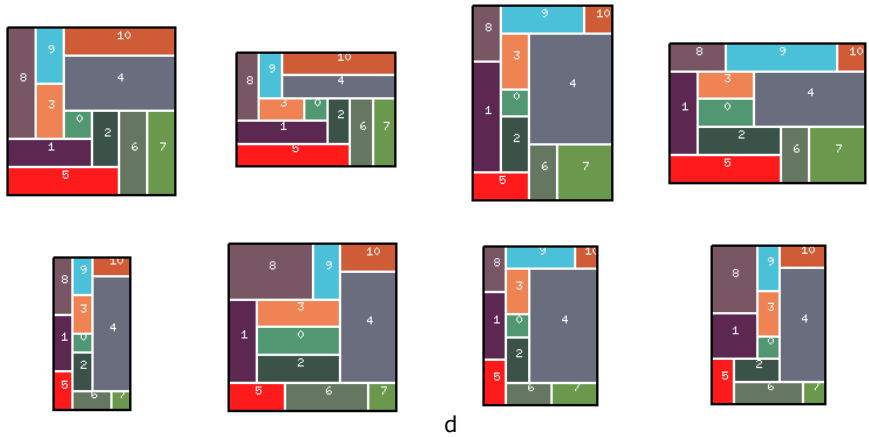

**Fig. 26.** a) Villa Trissino floorplan designed by Ottavio [48] b) Underlying adjacency graph of Villa Trissino taken as input by GPLAN c) Re-generated Villa Trissino floorplan by GPLAN d) Some floorplans that are topological distinct with Villa Trissino floorplan

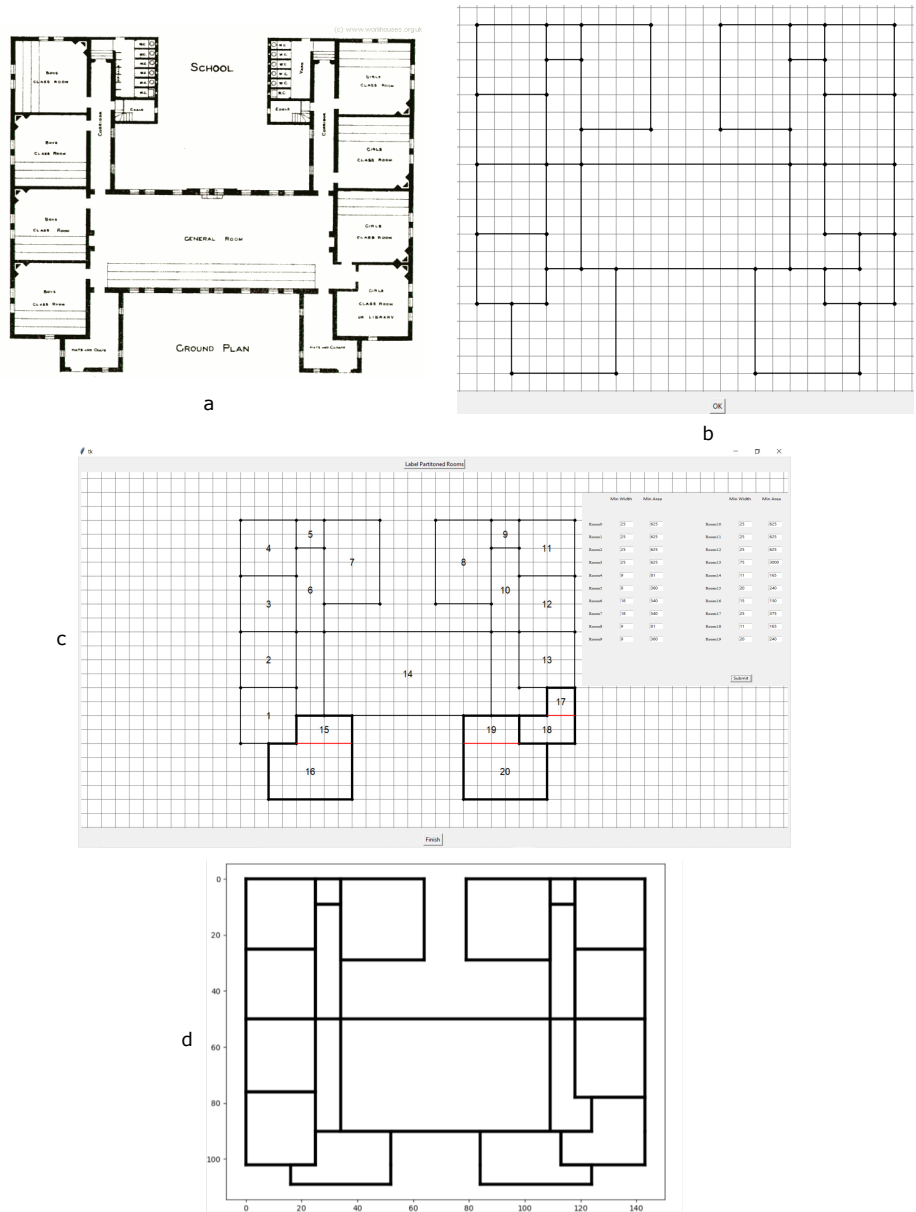

**Fig. 27.** a) Banstead Home School Plan by Higginbotham [49] b) Drawing the Banstead Home School Plan on a GUI generated by GPLAN c) Partitioning of rectilinear rooms and dimensional constraints d) Re-generated Banstead Home School Plan (with new dimensions) by GPLAN

# 6 Conclusion and Limitations

This work presents the automated generation of floorplans based on the following two cases:

i. Floorplans with rectangular boundary: For a given PTG $G$, there always exists a floorplan, and there exist many algorithms for generating either an RFP or an OFP corresponding to $G$. In this paper, instead of considering RFP and OFP separately, we proposed an approach which first generates an RFP if it exists; otherwise, it generates an OFP. By doing this, the user has more flexibility in considering the adjacency relations, i.e., it is not required for the user to have prior knowledge of graphs for which an RFP does not exists and the work is not limited to a specific class of graphs. Furthermore, for a given set of adjacencies, the user has a lot of choices in terms of topologically distinct layouts. The additional feature of GPLAN is its ability to generate a feasible dimensioned layout for any given dimensions. Furthermore, GPLAN is very efficient in handling the graphs with a large number of vertices, at the same time, it can quickly generate a large number of layouts for the given adjacencies.

ii. Floor plans with non-rectangular boundary: These floorplans are comparatively difficult to handle because of the flexibility in the boundary layout and this is why there does not exist any promising work for the automated generation of IFP corresponding to the given graphs. At the same time, introducing dimensions to these layouts is also challenging as compared to RFPs and OFPs. Hence, the proposed work can be seen as an alternative approach for building dimensioned IFP where the idea is to insert the adjacency relation through dimensionless layouts which, by default, also considers the geometry of rooms. Then, dimensioned IFPs are produced while preserving the given adjacencies, positions and shapes of the drawn rooms.

As mentioned in the Section 1, GPLAN can be seen as a beneficial tool for architects/designers, which is capable of generating a set of dimensioned floorplans for given adjacency relations; at the same time, it can also be used to re-generate existing floorplans. Although GPLAN has its merits, but it has the following limitations which we need to address in the near future:

i. It can be seen in Section 3.5 that GPLAN is capable of producing a very large number of solutions but it is not feasible for the user/designer to go through each solution. Hence, we need to identify and pick good architectural layouts from the obtained solution set. The first step in this direction is to restrict solutions on the basis of boundary constraints, i.e., for each room adjacent to the exterior, the user will be asked to choose its preferred location based on cardinal and inter-cardinal directions. It is possible to incorporate this part to GPLAN in future because GPLAN finds all possible boundary solutions for a given adjacency graph (as discussed in 3.5) to generate topologically distinct floorplans. For example, Figure 28a shows an adjacency graph with 3 vertices and Figure 28b shows the boundary

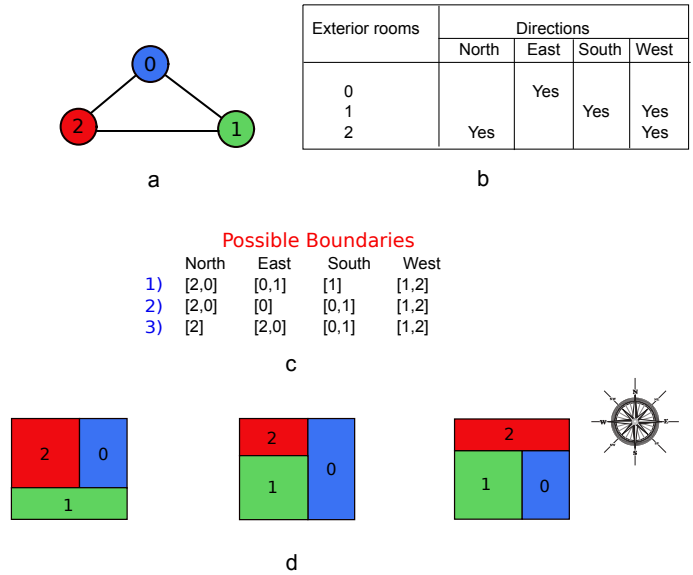

**Fig. 28.** a) An adjacency graph b) User-defined constraints c) Possible boundaries based on user-defined constraints d) Floor plans restricted to user-defined boundary constraints

    constraints defined by the user. Out of the 12 possible boundaries, 3 boundaries satisfy the user defined constraints as shown in Figure 28c, and their corresponding floorplans are shown in Figure 28d. The solutions satisfying user-defined boundary constraints will be further sorted on the basis of daylight and other architectural constraints which we need to collect as feedback from the architects while presenting GPLAN to them. Other than the floor plan assessment, we are also planning to introduce circulations to the floorplans obtained using GPLAN. A graph-theoretical approach for inserting circulations is given by Baybars [50] and there are some recent works in this direction, for example [51]. By exploring all available possibilities and understanding the architectural requirements, we will try to generate floorplans with circulations using GPLAN. At large, our objective is to adapt GPLAN for residential buildings while considering functionality and other architectural inputs.

ii. Since IFPs are more suitable for complex building structures like hospitals and universities, a separate study is required for the automated generation of IFPs for the given adjacencies. One of the ideas is to consider the distance matrix along with the adjacency graph for constructing dimensioned IFPs while considering the boundary layout as input.

    We acknowledge that architectural design is a multi-disciplinary and multi-constraints problem where producing an optimum solution which satisfies all

constraints and is simultaneously acceptable to architects is near to impossible. Therefore, computers cannot be a replacement to architects; nevertheless, at the same time, they can provide a variety of good initial layouts. Hence, in this paper, we presented GPLAN, which can be seen as a major contribution towards the automated generation of floorplans and it can be taken to new heights after getting inputs from designers/architects.

## Acknowledgement

The research described in this paper evolved as part of the research project Mathematics-aided Architectural Design Layouts (File Number: ECR/2017/000356) funded by the Science and Engineering Research Board, India.